\documentclass[aps,prl,twocolumn,nopacs,floatfix,amsmath,nofootinbib,amssymb,floatfix]{revtex4}
\usepackage{epsfig}
\usepackage{amsmath}
\usepackage{graphicx}
\usepackage{tikz}
\usepackage{tikz-feynman}
\usepackage{multirow}
\usepackage{hyperref}
\usetikzlibrary{arrows.meta, matrix}

\begin{document}


\title{Proper constituent gluon mass as the final piece to construct hybrid}

\author{Zi-Xuan Ma$^1$}
\author{Qi Huang$^1$}\email{06289@njnu.edu.cn}
\author{Rui Chen$^{2,3}$}\email{chenrui@hunnu.edu.cn}
\author{Li-Ming Wang$^4$}\email{lmwang@ysu.edu.cn}
\author{Yue Tan$^5$}\email{181001003@njnu.edu.cn}
\author{Xiao-Huang Hu$^6$}\email{201001002@njnu.edu.cn}
\author{Jun He$^1$}
\author{Hong-Xia Huang$^1$}

\affiliation{$^1$Department of Physics and Technology, Nanjing Normal University, Nanjing 210023, China\\
$^2$Hunan Research Center of the Basic Discipline for Quantum Effects and Quantum Technologies, Hunan Normal University, Changsha 410081, China\\
$^3$Key Laboratory of Low-Dimensional Quantum Structures and Quantum Control of Ministry of Education, Department of Physics and Synergetic Innovation Center for Quantum Effects and Applications, Hunan Normal University, Changsha 410081, China\\
$^4$Key Laboratory for Microstructural Material Physics of Hebei Province, School of Science, Yanshan University, Qinhuangdao 066004, China\\
$^5$Department of Physics, Yancheng Institute of Technology, Yancheng 224000, China\\
$^6$Department of Physics, Changzhou Institute of Industry Technology, Changzhou 213164, China}

\begin{abstract}

In this letter, we propose that a proper constituent gluon mass $m_g$=450 MeV can be applied to identify the hybrids composed of quarks and gluons. By investigating the spectra and decay widths of the light hybrids $(q\bar{q}g)$ with $J^P=1^{-+}$, we find the $\pi_1(1600)$ and $\eta_1(1855)$ may not be explained as $1^{-+}$ hybrids, simultaneously, and the $\eta_1(1855)$ observed by BESIII may not be a hybrid. In addition, we predict an existence of a hybrid $\eta_1(1640)$, which can be verified by searching the $a_1(1260)\pi$ channel. Moreover, we suggest the $K_1(1270)\bar{K}$ and $K_1(1270)\pi$ as the golden channels to search for an isospin-0 and an isospin-$\frac{1}{2}$ hybrids, respectively.

\end{abstract}

\maketitle

\section{Introduction}

As the fundamental theory of strong interaction, the Quantum Chromodynamics (QCD) reveals us all the possible interactions between quarks, anti-quarks, and gluons. Thus, theoretically, it can present us explanations on nearly all the phenomena in strong interaction field. However, due to the complex mathematical structure of this theory, direct calculations that started from the first principle are usually too difficult to carry on. As a result, various phenomenological models obeying the basic requirements of QCD are successfully proposed, which not only stand for our understandings on the strong interactions, but also were adopted to reveal the inner structures of the new hadronic structures (see review papers \cite{Chen:2016qju, Liu:2019zoy, Chen:2016spr, Guo:2017jvc, Chen:2022asf, Liu:2013waa, Hosaka:2016pey,Meng:2022ozq} for more details).

Among these QCD based phenomenological models, the potential model is one of the most successful proposals, where the quark model, one-gluon exchange potential that directly derived from QCD Lagrangian, and a phenomenological description on color confinement are combined together. After considering different modifications such as Goldstone boson exchange \cite{Vijande:2004he,Yang:2011rp,Chen:2024ukv,Yang:2017rmm,Huang:2015uda}, hidden local symmetry \cite{He:2023ucd,He:2023gqh}, scalar meson exchange \cite{Yang:2017qan,Vijande:2009pu,Tan:2022pzi}, semi-relativistic or relativistic \cite{Oettel:2000uw,Aslanzadeh:2017hhz,Godfrey:1985xj,Capstick:1986ter}, unquenched effects \cite{Tan:2019qwe,Ni:2025gvx,Ni:2023lvx,Santopinto:2010zza,Bijker:2012zza,Cardoso:2014xda}, etc., a global description on the spectra of traditional hadrons (mesons and baryons) can be actually well obtained. Especially, for the low-lying states, their masses can be very nicely reproduced \cite{Vijande:2004he,Yang:2011rp,Chen:2024ukv,He:2023gqh}.

However, with the rapid development of experiments, there appears a large number of hadronic states that cannot be contained into the traditional quark model, although their existences are permitted by QCD. Among these states, one kind of the convincing ones might be those with exotic quantum numbers. For example, three $J^{PC}=1^{-+}$ states $\pi_1(1400/1600)$ \cite{E852:1998mbq,CrystalBarrel:2019zqh,JPAC:2018zyd,Kopf:2020yoa,IHEP-Brussels-LosAlamos-AnnecyLAPP:1988iqi,Aoyagi:1993kn,E852:1997gvf,VES:2001rwn,CrystalBarrel:1998cfz,Meyer:2010ku,Baker:2003jh,Khokhlov:2000tk,COMPASS:2009xrl,CLEO:2011upl}, $\pi_1(2015)$ \cite{E852:2004gpn,E852:2004rfa}, and a new iso-scalar state $\eta_1(1855)$ that was observed by the BESIII Collaboration in the $J/\psi \to \gamma\eta\eta^\prime$ process \cite{BESIII:2022riz,BESIII:2022iwi}. The classification of them naturally becomes an issue in the community. Currently, there are two main opinions for their interpretations, one is the tetraquark state \cite{Chen:2008qw,Wan:2022xkx,Dong:2022cuw,Yang:2022rck,Yan:2023vbh}, the other is hybrid \cite{Narison:1999hg,Shastry:2022mhk,Chen:2023ukh,Iddir:1988jd,Benhamida:2019nfx,Eshraim:2020ucw,Zhang:2025xee,Dudek:2010wm,Qiu:2022ktc,Chen:2022qpd}, or a mixture of tetraquark and hybrid \cite{Narison:2009vj}. 

Obviously, from the view of constituent quark model, to build and complete the hybird mass spectra, the mass of the constituent gluon is the key factor. Actually, studies on the effective mass of gluon had already been carried out \cite{Binosi:2012sj,Binosi:2022djx,Ding:2022ows,Pelaez:2021tpq,Pelaez:2022rwx,Pelaez:2017bhh,vanEgmond:2021jyx,Reinosa:2014ooa,Cornwall:1981zr,Cornwall:1982zn,Cornwall:1989gv,Halzen:1992vd,McNeile:1998cp,Mei:2002ip,Bernard:2003jd,Bernard:1981pg,MILC:1997usn,Lacock:1996ny,Barnes:1982zs,Barnes:1982tx,Donoghue:1983fy,Horn:1977rq,Ishida:1991mx,Szczepaniak:2001rg,Iddir:2007dq}. While in Refs~\cite{Binosi:2012sj,Binosi:2022djx,Ding:2022ows}, authors proposed that the gluons can be massive via the Schwinger mechanism, i.e., the non-perturbative effect of the gauge field will bring a correction into the gluon propagator and resulting into a mass term, and they gave that $m_g\approx$450 MeV. Thus, in this letter, after adding the constituent gluon mass into the potential model and taking the same parameters given by meson spectra \cite{Vijande:2004he,Yang:2011rp,Chen:2024ukv}, we recalculate the spectra of low-lying hybrids in the light sector with quantum number $J^{PC}=1^{-+}$ by treating hybrid as a three-body system, and the results show a very consistent behavior as previous theoretical works, especially for the ground states \cite{Bernard:2003jd,Dudek:2013yja,Tan:2024grd,Meyer:2015eta,Shastry:2022mhk}. Furthermore, we calculate their two body strong decays at leading order, and the results can also match the corresponding ones in Ref.~\cite{Chen:2023ukh,Zhang:2025xee}, which show that a proper constituent gluon mass may be the final piece to construct hybrid.

\section{Model Setup}

\begin{figure}[!htbp]
    \centering
    \includegraphics[width=0.38\textwidth]{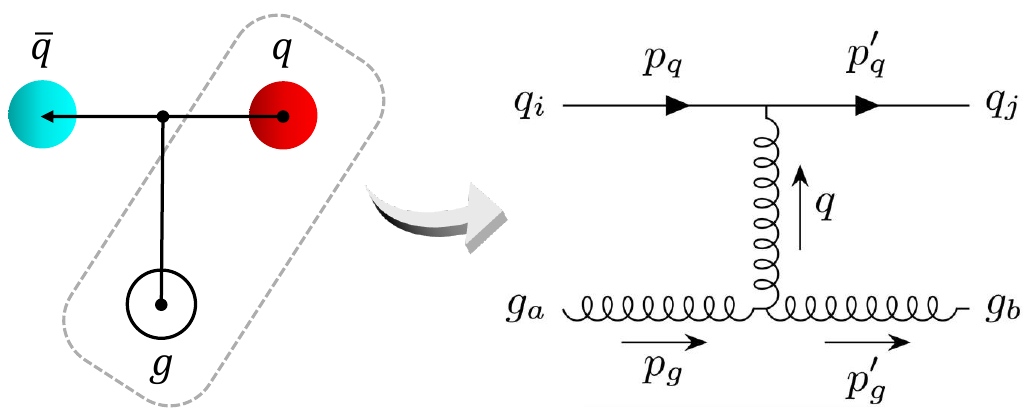}
    \caption{The coupling scheme of hybrid (left) and the interaction between quark and gluon (right). Here, in left panel, the T-type solid lines means that quark and antiquark couple first to form a $(q\bar{q})$ cluster, then gluon couples to this cluster to form the final hybrid.}
    \label{fig:qg}
\end{figure}

After taking the gluon as a constituent, our treatment on the hybrid then becomes a three-body problem, whose coupling scheme can be given by Fig.~\ref{fig:qg}.

We adopt the Gauss Expansion Method \cite{Hiyama:2003cu} for the spectra calculations, which is often used to study few-body systems. The relevant Hamiltonian can be constructed as
\begin{eqnarray}
    \hat{H} & =& \frac{\hat{p}_{q,\bar{q}}^{2}}{2\mu_{q,\bar{q}}} + \frac{\hat{p}_{q\bar{q},g}^{2}}{2\mu_{q\bar{q},g}} + \hat{V}_{q\bar{q}} + \hat{V}_{qg} + \hat{V}_{\bar{q}g},
\end{eqnarray}
with $\hat{p}_{a,b}$ and $\mu_{a,b}$ being the relative momentum and reduced mass between $a$ and $b$ clusters,  respectively. For the interaction between $q\bar{q}$, the potential keeps the same as chiral constituent quark model \cite{Vijande:2004he,Yang:2017rmm,Huang:2015uda}, i.e., 
\begin{eqnarray}
    V_{q\bar{q}}(\boldsymbol{r}) = V_{q\bar{q}}^{CON}(\boldsymbol{r}) + V_{q\bar{q}}^{OGE}(\boldsymbol{r}) + V_{q\bar{q}}^{GBE}(\boldsymbol{r}),
\end{eqnarray}
where $V_{q\bar{q}}^{CON}$, $V_{q\bar{q}}^{OGE}$, and $V_{q\bar{q}}^{GBE}$ denote the confinement, one gluon exchange, and Goldstone boson exchange, respectively \cite{Vijande:2004he,Yang:2017rmm,Huang:2015uda}. While the interaction between (anti-)quark and gluon contains the confinement and one gluon exchange (right panel in Fig.~\ref{fig:qg}) only as
\begin{eqnarray}
    V_{qg}(\boldsymbol{r}) &=& V_{qg}^{CON}(\boldsymbol{r}) + V_{qg}^{OGE}(\boldsymbol{r}),\\
    V_{qg}^{CON}(\boldsymbol{r}) &=& V_{q\bar{q}}^{CON}(\boldsymbol{r})~\mathrm{with}~\boldsymbol{\lambda}_c \cdot \boldsymbol{\lambda}_c^\ast \rightarrow i\boldsymbol{\lambda}^d \cdot \boldsymbol{f}^d,\\
    V_{\bar{q}g}(\boldsymbol{r}) &=& V_{qg}(\boldsymbol{r})~\mathrm{with}~\boldsymbol{\lambda}_c \rightarrow -\boldsymbol{\lambda}_c^\ast,
\end{eqnarray}
where $\boldsymbol{\lambda}$ denotes the Gell-Man matrices, $\boldsymbol{f}$ are eight $8 \times 8$ matrices, whose matrix elements are just the anti-symmetric structure constants of SU(3) group, i.e., $(f_c)_{ab}=f_{cab}$.  $V_{qg}^{OGE}(\boldsymbol{r})$ is derived directly from QCD Lagrangian. By adopting the non-relativistic reduction in addition with Fourier transformation on the $t$-channel scattering amplitude between quark and gluon (right panel in Fig.~\ref{fig:qg}) \cite{Mathieu:2007fpv,Varshalovich:1988ifq}, its expression can be explicitly written in coordinate representation as
\begin{eqnarray*}
    V_{qg}(\boldsymbol{r}) &=& \frac{\alpha_s}{2}  \boldsymbol{\lambda}_c \cdot \boldsymbol{f}_c \left[\frac{1}{r}-\left(\frac{2\pi}{3m_g^2}+\frac{\pi}{2m_q^2}\right)\delta(\boldsymbol{r})\right.\nonumber\\
    &&+ \frac{1}{2m_g^2r^3}\left(\boldsymbol{s}_g \cdot (\boldsymbol{l}_g+ \boldsymbol{s}_g)-3\frac{(\boldsymbol{s}_g \cdot \boldsymbol{r})(\boldsymbol{s}_g \cdot \boldsymbol{r})}{r^2}\right) \nonumber\\
	&&\left.-\frac{\boldsymbol{s}_q \cdot \boldsymbol{l}_q}{2m_q^2r^3}-\frac{\boldsymbol{s}_g \cdot \boldsymbol{l}_q-\boldsymbol{s}_q \cdot \boldsymbol{l}_g}{m_gm_qr^3}
 -\frac{8\pi \boldsymbol{s}_g \cdot \boldsymbol{s}_q\delta(\boldsymbol{r})}{3m_gm_q}\right],
\end{eqnarray*}
with $\alpha_s$ being the effective scale-dependent running coupling constant taking the famous form as in Refs. \cite{Vijande:2004he}, $\alpha_{s}(\mu) =\alpha_{0}/{\log\left({\mu^2+\mu_{0}^2}/{\Lambda_{0}^2}\right)}$,
where $\mu$ is the reduced mass of two constituents, and the Dirac function is smeared as \cite{Vijande:2004he,Yang:2011rp,Chen:2024ukv,Yang:2017rmm,Huang:2015uda}
$\delta(\boldsymbol{r}) \to {\mu e^{-\mu r/r_0}}/({4\pi r_0 r})$.

Next, for the wave function of $q\bar{q}g$ hybrid, to keep the same with Refs.~\cite{Chen:2023ukh,Zhang:2025xee,Ding:2006ya}, its construction procedure is explicitly written as
\begin{eqnarray}
    \psi_{q\bar{q}g}^J &=& \left[ \left[\psi_q^{\frac{1}{2}}\psi_{\bar{q}}^{\frac{1}{2}}\right]^{S_{q\bar{q}}} \left[ \left[\psi_{q\bar{q},g}^{L_{q\bar{q},g}} \psi_g^1\right]^{J_g} \psi_{q\bar{q}}^{L_{q\bar{q}}} \right]^{L_g} \right]^J \nonumber\\
    && \otimes \psi_{q\bar{q}g}^c \otimes \psi_{q\bar{q}g}^f.\label{eq:gluoncloud}
\end{eqnarray}
Here, $\psi_{q\bar{q}g}^f$ is the flavor wave function, that is nearly the same as corresponding meson but inserting into one more gluon $g$. While $\psi_{q\bar{q}g}^c$ is the color wave function as \cite{Mathieu:2007fpv}
$| \psi_{q\bar{q}g}^c \rangle= \frac{\delta_{ad}}{\sqrt{8}} \frac{(\lambda^a)_{bc}}{\sqrt{2}} f^{dbc} ~ |q^b\rangle \otimes |\bar{q}^c\rangle \otimes |g^d\rangle$.
For the orbital wave functions $\psi_{q\bar{q},g}^{L_{q\bar{q},g}}$ and $\psi_{q\bar{q}}^{L_{q\bar{q}}}$, they are expanded in coordinate representation by a series of Gaussian basis as
$\psi_i^{L_i}(\boldsymbol{r}) = \sum_{n=1}^{n_{max}} c_n^{L_i} N_{n L_i} e^{-\nu_n r^2} \mathcal{Y}_{L_i}(\boldsymbol{r})$,
where $c_n^{L_i}$ are the coefficients to be determined by variational principle, $\mathcal{Y}_{L_i}(\boldsymbol{r})$ denotes the solid spherical harmonics, $N_{n L_i}$ is the normalization factor written as
$N_{n L_i} = \sqrt{{2^{L_i+2}(2\nu_n)^{L_i+3/2}}/({\sqrt{\pi}(2L_i+1)!!})}$,
and $\nu_n=1/r_n^2$ is parameterized with $r_0$ and $r_{max}$ with
$r_n=r_0 \left(\frac{r_{max}}{r_0}\right)^{\frac{n-1}{n_{max}-1}}$. In this work, to be the same with meson spectra calculations \cite{Vijande:2004he,Yang:2011rp,Chen:2024ukv}, we take $r_0$=0.1 fm, $r_{max}$=2 fm, and $n_{max}$=8.

Then, for the strong decay of $q\bar{q}g$ hybrid, the simplest case is that the quark-antiquark pair generated from the constituent gluon combines with the remaining $q\bar{q}$ to form the two mesons. Thus, the Hamiltonian is
\begin{eqnarray}
    \hat{H}_I = i \sqrt{4\pi\alpha_s} \frac{(\lambda^a)_{bc}}{2} \int d^3 \vec{x} ~ \bar{q}^c(\vec{x}) \gamma^\mu q^b(\vec{x}) A_\mu^a(\vec{x}),
\end{eqnarray}
with $q^b$, $\bar{q}^c$, and $A_\mu^a$ being the quark, anti-quark, and gluon fields, respectively. Then, by field expansions and non-relativistic reduction \cite{Varshalovich:1988ifq,Ding:2006ya}, the leading order transition operator can be further rewritten as
\begin{eqnarray}
    \hat{T} &=& 3 i \sqrt{\pi\alpha_s} (\lambda^a)_{bc} \sum_{s,s^\prime,m} \int \frac{d^3 \vec{p}_1 d^3 \vec{p}_2 d^3 \vec{k}}{\sqrt{2m_g}(2\pi)^6} \delta(\vec{p}_1+\vec{p}_2-\vec{k})\nonumber\\
    &&\times C_{1,m;1,-m}^{0,0}C_{1,-m;1/2,s^{\prime}}^{1/2,s}d_{s^\prime}^{c\dagger}(\vec{p}_1) b_s^{b\dagger}(\vec{p}_2) a_m^a(\vec{k}),
\end{eqnarray}
where $a_m^a$ is the annihilation operator of gluon, and $b_s^{b\dagger}$ and $d_{s^\prime}^{c\dagger}$ are the creation operators of quark and anti-quark, respectively. Once acting it onto the wave functions of initial hybrid $A$ and final mesons $B$ and $C$ to obtain the helicity amplitude $\mathcal{M}_{A \to BC}^{spin}$, the decay width can be finally calculated as
\begin{eqnarray}
    \Gamma_{A \to B C} = \frac{1}{1+\delta_{BC}}\frac{p_B E_B E_c}{\pi M_A} \sum_{spin}\frac{|\mathcal{M}_{A \to BC}^{spin}|^2}{2J_A+1}.
\end{eqnarray}
One can see the appendix for a detailed introduction to the interactions and wave functions.

\section{Numerical results and discussions}

We use the same quantum number configuration as our previous work \cite{Zhang:2025xee} to present our numerical results, where to form a $1^{-+}$ hybrid with lowest mass, the constituent gluon should be a transverse electric gluon, and the hybrid should be a gluon-excited state. Translating such configuration into the quantum numbers used in Eq.~(\ref{eq:gluoncloud}), it means $S_{q\bar{q}}=1$, $L_{q\bar{q}}=0$, $L_{q\bar{q},g}=1$, $J_g=1$, $L_g=1$, and $J=1$.

\begin{table}[!h]
    \centering
    \caption{Parameters of the chiral quark model with three confinement potential forms: screened(n=0) \cite{Vijande:2004he}, linear(n=1) \cite{Yang:2011rp}, square(n=2) \cite{Chen:2024ukv}. The Goldstone-boson exchange interaction parameters are consistent. Masses of $\pi$, $\eta$, $K$ adopt experimental values, while other parameters --- $m_\sigma$ = 3.42 $\rm fm^{-1}$, $\Lambda_\pi = \Lambda_\sigma$ = 4.2 $\rm fm^{-1}$, $\Lambda_\eta = \Lambda_K$ = 5.2 $\rm fm^{-1}$, $\theta_{p} = -15^\circ$, $g_{ch}^2/(4\pi)$ = 0.54 --- are adopted from Ref.~\cite{Vijande:2004he}.}\label{tab:parameter}
    \renewcommand\arraystretch{1.4}
    \setlength{\tabcolsep}{1.5mm}     
    \begin{tabular}{cccc} \hline\hline   
    & & & [~~Scr.~,~~Lin.~~,~Squ.~~]\\\hline 

    Gluon~mass & $m_{g}$~(MeV) & & 450\\\hline
    
    Quark~masses & $m_{u,d}$~(MeV) & & 313\\
    
    & $m_{s}$~(MeV) & & [~~555~~,~~525~~,~~536~~]\\
    \hline
    
    Confinement & $a_{c}$~$\rm (MeV\cdot fm^{-n})$ & & [~~430~~,~~160~~,~~101~~]\\
    
    & $\Delta$~(MeV) & & [~181.1,~-131.1,~-78.3~]\\
    
    & $\mu _{c}$~$\rm (fm^{-1})$ & & [~~0.7~~,~~~~-~~~~,~~~~-~~~]\\
    
    & $a _{s} $ & & 0.777\\
    
    \hline
    
    OGE & $\alpha_{0}$ & & [~2.12~,~~2.65~~,~3.67~]\\
    
    & $\Lambda_{0}$~$\rm (fm^{-1})$ & & [~0.113,~~0.075,~0.033]\\
    
    & $\mu_{0}$~(MeV) & & 36.976\\
    
    & $\hat{r}_{0}$~(MeV$\cdot$fm) & & 28.17\\
    
    & $\hat{r}_{g}$~(MeV$\cdot$fm) & & 34.5\\
    
    \hline\hline
    \end{tabular}
\end{table}

Then, we adopt three kinds of confinements to make our calculation, which are in screened form, linear form, and square form, respectively. These confinements actually represent different estimations on the unquenched effect~\cite{Chen:2017mug,Li:2009ad}. However, different confinements should have the least impact on the ground state but have considerable effect on the excited one. Here, the adopted model parameters are summarized in Table~\ref{tab:parameter}, which keeps the same values as given in Refs.~\cite{Vijande:2004he,Yang:2011rp,Chen:2024ukv}, but just adds into the constituent gluon mass $m_g$=450 MeV as the final piece, our spectra on low-lying light hybrids are presented by the black lines in Fig.~\ref{fig:qg}.

\begin{figure}[!h]
    \centering
    \includegraphics[width=0.45\textwidth]{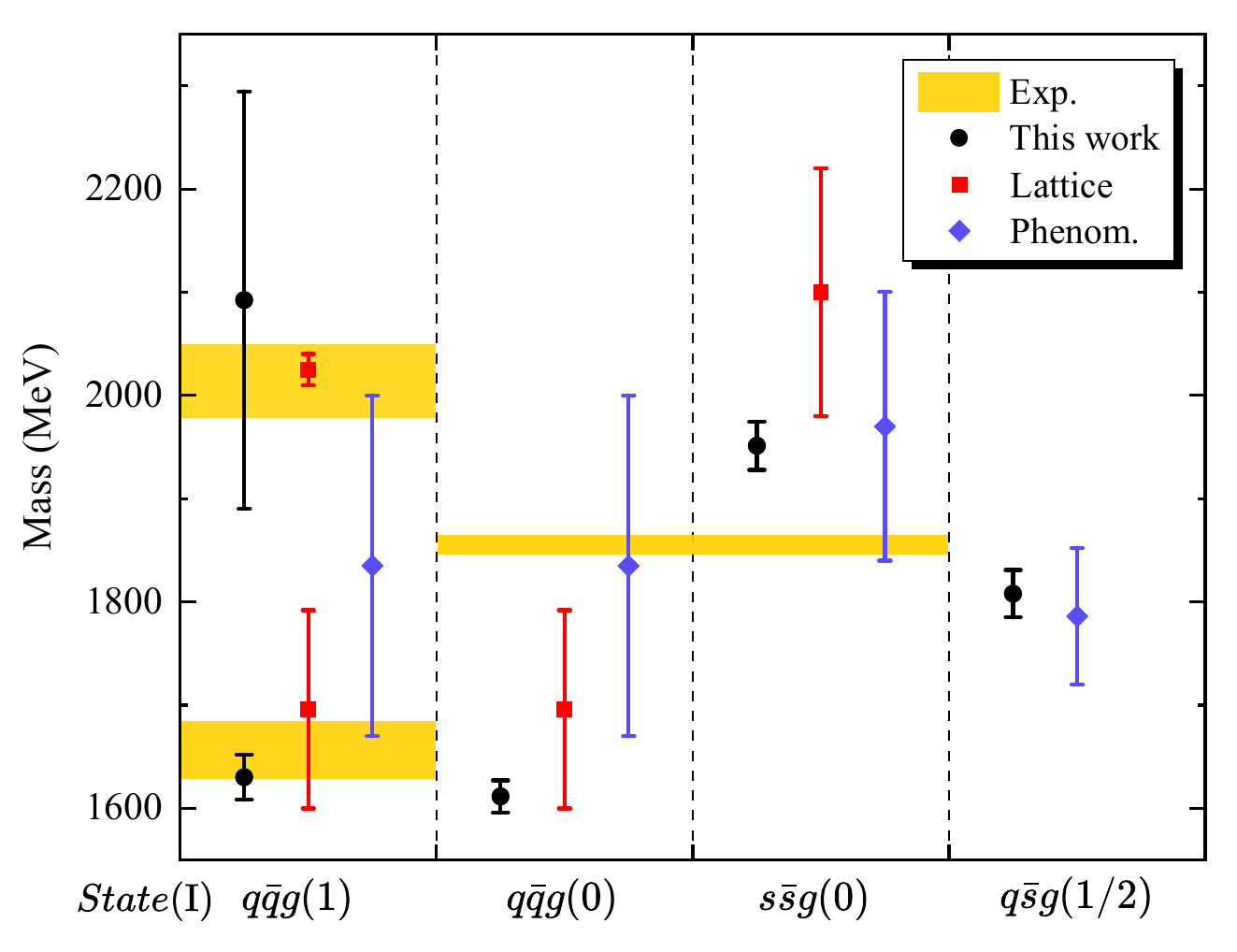}
    \caption{Mass spectra of low-lying hybrid states with $I(J^{PC})=I(1^{-+})$, where the error bar line means a collection of different results obtained by different models or works. Here, the black lines, yellow bands, red lines, and blue lines correspond to the results in our work, the experimental data from Ref.~\cite{ParticleDataGroup:2024cfk}, the lattice QCD results referenced in \cite{Bernard:1981pg,Lacock:1996ny,MILC:1997usn,McNeile:1998cp,Mei:2002ip,Bernard:2003jd,Dudek:2013yja}, and in other phenomenological models \cite{Meyer:2015eta,Shastry:2022mhk,Barnes:1982zs,Barnes:1982tx,Donoghue:1983fy,Iddir:2007dq,Szczepaniak:2001rg,Ishida:1991mx,Horn:1977rq,Tan:2024grd}, respectively.} 
    \label{Mass}
\end{figure}

As we can see from Fig.~\ref{fig:qg}, for ground states, our results can match very well with lattice QCD and other phenomenological models. Especially, for the ground $q\bar{q}g$ state with isospin-1, its position is just located at the experimental result of $\pi_1(1600)$. While for the ground $q\bar{q}g$ and $s\bar{s}g$ states with isospin-0, their masses can also match with other theoretical results, although they are a little far away from the experimental results of $\eta_1(1855)$. Actually, such phenomena are natural, since the physical isospin-0 hybrid in the light region is finally a mixture of $q\bar{q}g$ and $s\bar{s}g$ as
\begin{eqnarray}
    \begin{bmatrix} \eta_1^{low} \\ \eta_1^{high} \end{bmatrix} = \begin{bmatrix} \cos\theta & -\sin\theta \\ \sin\theta & \cos\theta \end{bmatrix} \begin{bmatrix} q\bar{q}g \\ s\bar{s}g \end{bmatrix}.
\end{eqnarray}
Thus, there should exist two states $\eta_1^{low}$ and $\eta_1^{high}$, whose masses are located between the masses of pure $q\bar{q}g$ and $s\bar{s}g$ hybrids. If we take $\eta_1^{high}$ as $\eta_1(1855)$, as in our previous work \cite{Zhang:2025xee}, we can determine the mixing angle to be about $28.8^{\circ}$, with which the location of $\eta_1^{low}$ state is then around 1640 MeV, below we will denote it as $\eta_1(1640)$. Finally, for the ground $q\bar{s}g$ state, we predict that it is located around 1.8 GeV, which is consistent with Ref.~\cite{Chen:2023ukh}.

However, for the first excited state of $q\bar{q}g$ hybrid with isospin-1, our result exhibits a very large uncertainty. That is because when treating the hybrid as a three-body problem, there are two possible exciting modes, one is between $q\bar{q}$, the other is between gluon and $q\bar{q}$ cluster. In addition, in our work, we adopt three kinds of confinements, which will have very different descriptions on the behaviors of excited states. As a result, the uncertainty of the first excited state of $q\bar{q}g$ hybrid with isospin-1 is very large, although it can contain the location of $\pi_1(2015)$.

Nevertheless, it may indeed show that QCD, and the potential model derived from it, is a uniform theory. That means as long as we can reproduce the spectra of mesons and baryons, since the coupling constant in QCD is global, a generalization to the spectra of hybrids, at least for ground states, may be convincing. According to our further calculations, this primary conclusion may also be valid for decay widths, and our results on widths are collected in Table~\ref{tab:groundwidth}

\begin{table}[ht]
    \centering
    \caption{Numerical results and comparisons of decay widths (in MeV$\times\alpha_s$ unit just as in Refs.~\cite{Chen:2023ukh,Zhang:2025xee,Ding:2006ya}) for $\pi_{1}(1600)$, $\eta_{1}(1640)$, $\eta_{1}^{\prime}(1855)$, and $K_{1}(1^{-+})$ in screened, linear and square confinements (abbreviated as [Scr., Lin., Squ.]). For the mixing angle of $\eta_{1}(1640)$ and $\eta_{1}(1855)$, as in Ref.~\cite{Zhang:2025xee}, we take $28.8^{\circ}$ for comparison. }\label{tab:groundwidth}
    \renewcommand\arraystretch{1.4}
    \begin{tabular}{cccc} \hline\hline
    States  & Channels &  [~Scr.~,~Lin.~,~Squ.~] & [~\cite{Chen:2023ukh}~,~\cite{Zhang:2025xee}~]  \\\hline
    $\pi_1(1600)$ & $b_1(1235)\pi$  &  [~53.3~,~69.7~,~88.0~]  & [~244~,~56.6~] \\
    & $f_1(1285)\pi$  &  [~~9.0~,~11.5~,~13.6~]  & [~~15~~,~~8.4~] \\
    & $\rho\pi$  &  [~~1.0~,~~1.3~~,~~1.4~~]  & [~~~2~~,~~~-~~~] \\\cline{2-4}
    & Total  &  [~63.3~,~82.5~,~103.0]  &  [~261~,~~65~~] \\\hline
    
    $\eta_1(1640)$ & $a_1(1260)\pi$ & [~35.5~,~45.8~,~58.3~] & [~~55~,~29.3~]\\
    & $\pi(1300)\pi$ & [~2.9~~,~~1.5~~,~~0.1~] & [~~~5~~,~~0.4~] \\\cline{2-4}
    & Total & [~38.4~,~47.3~,~58.4~] & [~~60~,~29.7~]\\\hline

    $\eta_{1}^\prime(1855)$ & $a_1(1260)\pi$ & [~11.0~,~16.6~,~24.5~] & [~~~-~~,~18.1~]\\
    & $f_1(1285)\eta$ & [~6.4~~,~~6.7~~,~~8.6~] & [~~~-~~,~~5.6~]\\
    &$\pi(1300)\pi$ & [~4.3~~,~~3.5~~,~~2.4~] & [~~~-~~,~~1.1~]\\
    &$K_1(1270)\bar{K}$ & [176.0~,~160.8~,~228.0] & [~157~,162.5]\\
    &$K^*\bar{K}$ & [~1.0~~,~~1.4~~,~~1.6~] & [~~~2~~,~~-~~~]\\\cline{2-4}
    &Total & [198.7~,~189.0~,~265.1] & [~159~,187.3]\\\hline

    $K_{1}(1^{-+})$&$K\pi$ & [~0.6~~,~~0.8~~,~~1.3~] & [~~~1~~,~~-~~~]\\
    &$K^*\pi$ & [~1.3~~,~~1.6~~,~~2.3~] & [~~~3~~,~~-~~~]\\
    &$K^*\eta$ & [~0.2~~,~~0.2~~,~~0.4~] & [~~~1~~,~~-~~~]\\
    &$K_1(1270)\pi$ & [~36.5~,~46.1~,~46.5~] & [~106~,~~-~~~]\\
    &$K_1(1400)\pi$ & [~0.0~~,~~0.0~~,~~2.2~] & [~146~,~~-~~~]\\
    &$h_1(1170)K$ & [~6.8~~,~~3.6~~,~~5.8~] & [~~16~,~~-~~~]\\
    &$K(1460)\pi$ & [~1.4~~,~~1.0~~,~~0.5~] & [~~~2~~,~~-~~~]\\\cline{2-4}
    &Total & [~46.8~,~53.3~,~59.0~] & [~275~,~~-~~~]\\\hline\hline
    \end{tabular}
\end{table}

As we can see from Table~\ref{tab:groundwidth}, although we treat the hybrid as a three-body problem and Refs.~\cite{Chen:2023ukh,Zhang:2025xee} treat it as a quasi two-body problem, and the model parameters we used are totally different. As if we get the similar spectra and adopt the same transition mechanism, most of our results are consistent to each other. Comparing to Ref.~\cite{Chen:2023ukh}, the main difference comes from the $b_1(1235)\pi$ channel of $\pi_1(1600)$, which results into the differences of $K_1(1270)\pi$ and $K_1(1400)\pi$ channels of $K_1(1^{-+})$, since $K_1(1270)$ and $K_1(1400)$ are mixtures of $K_1(^1P_1)$ and $K_1(^3P_1)$ states, and $b_1(1235)$ is also a $^1P_1$ state. While comparing to Ref.~\cite{Zhang:2025xee}, the main differences come from the $a_1(1260)\pi$ and $\pi(1300)\pi$ channels. According to our crosscheck, for $a_1(1260)\pi$ channel, it mainly due to the sensitivity on phase space caused by the node of Laguerre polynomials. While for $\pi(1300)\pi$ channel, it comes from the wave function of $\pi(1300)$, since in Ref.~\cite{Zhang:2025xee}, we use just one single Laguerre polynomial to simulate the wave function, which is less accurate than this work that the wave function of $\pi(1300)$ is also given by GEM method. 

As a result, according to our numerical calculations, we keep similar conclusions as in Ref.~\cite{Zhang:2025xee} that, in current decay mechanism, we cannot explain $\pi_1(1600)$ and $\eta_1(1855)$ as $1^{-+}$ hybrids simultaneously, and the $\eta_1(1855)$ observed by BESIII in $J/\psi \to \gamma \eta \eta^\prime$ may not be a hybrid since under this situation its decay width to $\eta \eta^\prime$ is nearly zero. Then, to search for an isospin-0 hybrid located around 1855 MeV and an isospin-$\frac{1}{2}$ hybrid located around 1800 MeV, the golden channels may be $K_1(1270)\bar{K}$ and $K_1(1270)\pi$, respectively. Furthermore, there may exist a partner of $\eta_1^\prime(1855)$, whose mass is around 1640 MeV, and future experiments can search the $a_1(1260)\pi$ channel to verify it. As for $\pi_1(1600)$, more precise analysis may be needed to see if it is really a broad structure.

In addition, as mentioned before, there already have many previous studies on the effective mass of a gluon. For example, in 1980s, John Cornwall had already demonstrated that gluons can acquire a gauge-invariant dynamical mass with their values around $500 \pm 200~\text{MeV}$~\cite{Cornwall:1981zr,Cornwall:1982zn,Cornwall:1989gv}. In Refs.~\cite{Barnes:1982zs,Barnes:1982tx,Donoghue:1983fy,Horn:1977rq,Ishida:1991mx,Szczepaniak:2001rg,Iddir:2007dq}, the effective gluon mass was given around 0.5-1.0~GeV by the bag model, lattice QCD and constituent gluon model. Recently, Refs.~\cite{Binosi:2012sj,Binosi:2022djx,Ding:2022ows} systematically reviewed their works on the emergence of hadron mass, and found an effective gluon mass on the order of $m_g \approx \frac{1}{2}m_p$, with $m_p$ being the mass of proton. Hence, we further use different gluon masses to study how they impact the spectra. The results show that, when the gluon mass is taken from 0.40 to 0.80 GeV, the mass positions of ground $1^{-+}$ hybrid mesons vary 1.58 to 1.90 GeV for the $q\bar{q}g$ state with $I=1$, 1.57 to 1.88 GeV for the $q\bar{q}g$ state with $I=0$, 1.91 to 2.19 GeV for the $s\bar{s}g$ state, 1.75 to 2.02 GeV for the $q\bar{s}g$ state, respectively. The masses of ground $1^{-+}$ hybrid mesons change less than the constituent gluon mass due to the nonlinear coupling constant $\alpha_s$. We also find our results still match the conclusions given by other models, which may prove our opinion in this work that proper constituent gluon mass is possibly the final piece to construct the hybrid meson spectra.

\section{Summary}

As a unified theory, QCD reveals us an image of strong interactions. Especially, the global property of the coupling constants in it makes us believe that, if we can reproduce some necessary parts of the physical phenomena in strong region, with the fixed parameters, maybe we can have a chance to get a whole picture.

As a trial, we take a look into the spectra of light hybrids with exotic quantum number $1^{-+}$, where the hybrid is treated as a three-body system and the gluon becomes a constituent. After taking the mass of the constituent gluon as $m_g \approx \frac{1}{2}m_p\approx$450 MeV \cite{Binosi:2012sj,Binosi:2022djx,Ding:2022ows}, we make a numerical calculation on the spectra at first. The results show that, as long as we add into the constituent gluon mass as the final piece, by just using the same model parameters obtained from meson spectra \cite{Vijande:2004he,Yang:2011rp,Chen:2024ukv}, we can immediately get the same conclusion on the light $1^{-+}$ hybrid spectra, especially for ground states \cite{Chen:2023ukh,Zhang:2025xee}.

Then, we calculate the partial decay widths of the ground light $1^{-+}$ hybrids with a transition operator coming from the quark-gluon vertex. It shows that almost all the results are consistent with Refs.~\cite{Chen:2023ukh,Zhang:2025xee}, which, in our view, also reflects the unification of potential model. Based on our numerical results, we find that we still cannot explain $\pi_1(1600)$ and $\eta_1(1855)$ as $1^{-+}$ hybrids simultaneously due to the total width, and the $\eta_1(1855)$ may not be a hybrid since its decay width to $\eta\eta^\prime$ at leading order is almost zero. In addition, $K_1(1270)\bar{K}$ and $K_1(1270)\pi$ channels may be the golden channels to search for the isospin-0 and isospin-$\frac{1}{2}$ hybrid respectively. Furthermore, there exists another $\eta_1(1640)$, and future experiments such as BESIII can search the $a_1(1260)\pi$ channel to verify it.

{\it Acknowledgments--}
The authors want to thank Ya-Qi Cui, Jia-Lun Ping, Fu-Yuan Zhang, and Hai-Qing Zhou for very useful discussions. This work is supported partly by the National Natural Science Foundation of China under Grant Nos. 12305087, 12305139, 12205249, 12475080, 12405104, and 11675080, the Start-up Funds of Nanjing Normal University under Grant No.~184080H201B20, the Natural Science Foundation of Hebei Province under Grant No.~A2022203026, the Higher Education Science and Technology Program of Hebei Province under Contract No.~BJK2024176, the Research and Cultivation Project of Yanshan University under Contract No.~2023LGQN010, the Funding for School-Level Research Projects of Yancheng Institute of Technology under Grant No. xjr2022039, the Programme of Natural Science Foundation of the Higher Education Institutions in Jiangsu under Contract No. 1020242167, and the Xiaoxiang Scholars Programme of Hunan Normal University.

\section*{Appendix}

\subsection{Some details on the model}

\subsubsection{Deriving the potential between quark and gluon}

To derive the potential between quark and gluon, We can start from the QCD Lagrangian, which is usually written as
\begin{eqnarray}
	\mathcal{L}_{\mathrm{QCD}} = \bar{\psi}_i (i\gamma^\mu D_\mu - m_i) \psi_i - \frac{1}{4} G^a_{\mu\nu} G^{a\mu\nu},
\end{eqnarray}
where $\psi_i$ is the quark field, $m_i$ is the quark mass, $D_\mu = \partial_\mu - ig_s G^a_\mu T^a$ is the covariant derivative, $G^a_{\mu\nu} = \partial_\mu G^a_\nu - \partial_\nu G^a_\mu + g_s f^{abc} G^b_\mu G^c_\nu$ is the gluon field strength tensor, $T^a$ is the generator of $SU(3)$ group, and $f^{abc}$ is the structure constant of $SU(3)$ group.

From this Lagrangian, the Feynman rules of 3-gluons-vertex and quark-gluon interaction can be derived, which can be expressed as follows,
\begin{widetext}
\begin{eqnarray}
	&&\begin{tikzpicture}[baseline=(o.base)]
		\node at (7,-1) {\( \begin{split} = - g_s f^{abc} \left[ (p_g^\prime-q)^\theta g^{\phi\mu} + (q+p_g)^\phi g^{\mu\theta}   - (p_g+p_g^\prime)^\mu g^{\theta \phi} \right] A_\theta^{b} A_\phi^{\dag c} A_\mu^{\dag a}, \end{split}\)};
        \begin{feynman}[inline=(o.base)]
			\vertex (o) at (0, 0);
            \vertex (a) [below = 1.5cm of o] { \( g_a \) };
            \vertex (b) [right = 1.5cm of o] { \( g_c \) };
            \vertex (c) [left = 1.5cm of o] { \( g_b \) };

            \diagram* {
                (a) -- [gluon,momentum={[arrow shorten=0.25mm,arrow distance=2.5mm]\(q,\mu\)}] (o) -- [gluon,rmomentum'={[arrow shorten=0.25mm,arrow distance=2.5mm]\(p_g^\prime,\phi\)}] (b),
                (o) -- [gluon,momentum={[arrow shorten=0.25mm,arrow distance=2.5mm]\(p_g,\theta\)}] (c),
            };
        \end{feynman}
    \end{tikzpicture}\nonumber\\ 
	&&\begin{tikzpicture}[baseline=(o.base)]
		\node at (3.6,-1) {\( \begin{split} = i g_s \bar{u}(p^\prime) \gamma^\mu u(p) T^a A_\mu^{\dag a}, \end{split}\)};
        \begin{feynman}[inline=(o.base)]
			\vertex (o) at (0, 0);
            \vertex (a) [below = 1.5cm of o] { \( g_a \) };
            \vertex (b) [right = 1.5cm of o] { \( q_j \) };
            \vertex (c) [left = 1.5cm of o] { \( q_i \) };

            \diagram* {
                (a) -- [gluon,rmomentum={[arrow shorten=0.25mm,arrow distance=2.5mm]\(q,\mu\)}] (o) -- [fermion,rmomentum'={[arrow shorten=0.25mm,arrow distance=1.0mm,arrow style=white,label style=black]\(p^\prime\)}] (b),
                (o) -- [anti fermion,momentum={[arrow shorten=0.25mm,arrow distance=1.0mm,arrow style=white, label style=black]\(p\)}] (c),
            };
        \end{feynman}
    \end{tikzpicture}
\end{eqnarray}
\end{widetext}
with $T^a=\frac{\lambda^a}{2}$, $A_\mu^a=\epsilon_\mu \otimes \phi_g^a$ is the gluon field in which $\epsilon_\mu$ and $\phi_g^a$ being the spin and color wave function of gluon respectively, $u(p)$ is the quark spinor, $g$ is the strong coupling constant, $f^{abc}$ is the structure constant of $SU(3)$ group, and $\lambda^a$ is the Gell-Mann matrix.

For convenience in calculating the spatial part of the potential, we define two structure functions as
\begin{eqnarray}
	G^\mu(q) &=& -\left[ (p_g^\prime-q)^\theta g^{\phi\mu} + (q+p_g)^\phi g^{\mu\theta}\right.\nonumber\\
    &&\left.- (p_g+p_g^\prime)^\mu g^{\theta \phi} \right] \epsilon_\theta(p) \epsilon^{~\dag}_{\phi}(p^\prime),\\
	J^\mu(q) &=& \bar{u}(p^\prime) \gamma^\mu u(p).
\end{eqnarray}
After non-relativistic reduction, both $G^\mu(q)$ and $J^\mu(q)$ can be divided into two parts as
\begin{eqnarray}
	J^0(\vec{q}) &=& 2m \chi_{\frac{1}{2}s^\prime}^\dag\left[1-\frac{\vec{q}^{~2}-2i\vec{\sigma} \cdot (\vec{q} \times \vec{p})}{8m^2}\right] \chi_{\frac{1}{2}s},\\
	\vec{J}(\vec{q}) &=& 2m \chi_{\frac{1}{2}s^\prime}^\dag \left[ \frac{i\vec{\sigma} \times \vec{q}}{2m}+\frac{\vec{q}+2\vec{p}}{2m} \right] \chi_{\frac{1}{2}s},\\
	G^0(\vec{q}) &=& -2m \chi_{1\sigma^\prime}^\dag\left[ 1+\frac{\vec{p}^{~2}+\vec{p}^{~\prime 2}}{4m^2} - \frac{i\left(\vec{q} \times \vec{p} \right)\cdot \vec{S}}{2m^2}\right.\nonumber\\
    &&\left.+\frac{\vec{q}^{~2}-(\vec{S} \cdot \vec{q})^2}{2m^2}\right] \chi_{1\sigma},\\
	\vec{G}(\vec{q}) &=&-(2\vec{p}-\vec{q})(\vec{\epsilon} \cdot \vec{\epsilon}^{~\dag})+2(\vec{q} \cdot \vec{\epsilon}^{~\dag})\vec{\epsilon}-2(\vec{q} \cdot \vec{\epsilon})\vec{\epsilon}^{~\dag},\nonumber\\
\end{eqnarray}
where $\chi_{s m_s}$ is the $m_s$ component of the spinor function with spin-$s$, $\vec{\epsilon}$ is the spin-1 polarization vector.

Then, the scattering amplitude of quark-gluon interaction is
\begin{eqnarray}
	i\mathcal{M}_{qg} &=& \left[\left(i \left(-G^\mu(q)\right)\right) \left( -g_{\mu\nu}+\frac{q^\mu q^\nu}{q^2} \right) \frac{i}{q^2} \left(i J^\nu(q)\right)\right]\nonumber\\
    &&\otimes \left[ -ig^2 \frac{1}{2} \boldsymbol{\lambda}^d \cdot \boldsymbol{f}^d \right],
\end{eqnarray}
by adopting Breit approximation in addition with $q=(0,\vec{q})$, the effective potential in momentum representation can be expressed as
\begin{eqnarray}
	V_{qg} = \frac{-\mathcal{M}_{qg}}{4 m_q m_g} \equiv \frac{U_{qg}}{{4 m_q m_g}} \otimes \left[ -ig^2 \frac{1}{2} \boldsymbol{\lambda}^d \cdot \boldsymbol{f}^d \right],
\end{eqnarray}
where $U_{qg}$ contains the spatial part only,
\begin{eqnarray}
	U_{qg} = \frac{\vec{G}(\vec{q}) \cdot \vec{J}(\vec{q})-G^0(\vec{q})J^0(\vec{q})}{\vec{q}^{~2}} - \frac{(\vec{q}\cdot\vec{G}(\vec{q}))(\vec{q}\cdot\vec{J}(\vec{q}))}{\vec{q}^{~4}}.\nonumber\\
\end{eqnarray}

With two properties of polarization vector \cite{Varshalovich:1988ifq},
\begin{eqnarray}
	(\vec{\epsilon}\cdot\vec{a})(\vec{\epsilon}^{~\dag}\cdot\vec{b}) &=& \chi_{1\sigma^\prime}^\dag \left[ \vec{b}\cdot\vec{a} + \frac{i}{2} (\vec{b} \times \vec{a}) \cdot \hat{S} - \frac{1}{2}(\hat{S}\cdot \vec{b})\right.\nonumber\\
    &&\left.\times(\hat{S} \cdot \vec{a})  - \frac{1}{2}(\hat{S}\cdot \vec{a})(\hat{S} \cdot \vec{b}) \right] \chi_{1\sigma},\\
	\vec{\epsilon}\cdot \vec{\epsilon}^{~\dag} &=& \chi_{1\sigma^\prime}^\dag \left[ \hat{I} \right] \chi_{1\sigma},
\end{eqnarray}
we can finally get the similar expression of $U$ as Ref.~\cite{Mathieu:2007fpv}
\begin{eqnarray}
	U&=&\frac{1}{\vec{q}^{~2}}-(\frac{1}{2m_g^2}+\frac{1}{8m_q^2})-\frac{\vec{S_q} \cdot \vec{S_g}}{m_gm_q} + \frac{i\vec{S_q} \cdot (\vec{q} \times \vec{p}_q)}{2m_q^2\vec{q}^{~2}}\nonumber\\
    &&-\frac{i\vec{S_g} \cdot (\vec{q} \times \vec{p}_g)}{2m_g^2\vec{q}^{~2}} +\frac{(\vec{S_g} \cdot \vec{q})^{~2}}{2m_g^2\vec{q}^{~2}} +\frac{(\vec{p}_g \cdot \vec{q})(\vec{p}_q \cdot \vec{q})}{m_gm_q\vec{q}^{~4}}\nonumber\\
    &&-\frac{\vec{p}_g \cdot \vec{p}_q}{m_gm_q\vec{q}^{~2}} +\frac{(\vec{S_q} \cdot \vec{q})(\vec{S_g} \cdot \vec{q})}{m_gm_q\vec{q}^{~2}}+\frac{i\vec{S_g} \cdot (\vec{q} \times \vec{p}_q)}{m_gm_q\vec{q}^{~2}}\nonumber\\
    &&-\frac{i\vec{S}_q \cdot (\vec{q} \times \vec{p}_g)}{m_gm_q\vec{q}^{~2}}.
\end{eqnarray}

\subsubsection{A little more on the construction of spin-orbit wave function}
\begin{figure}[!h]
    \centering
    \includegraphics[width=0.45\textwidth]{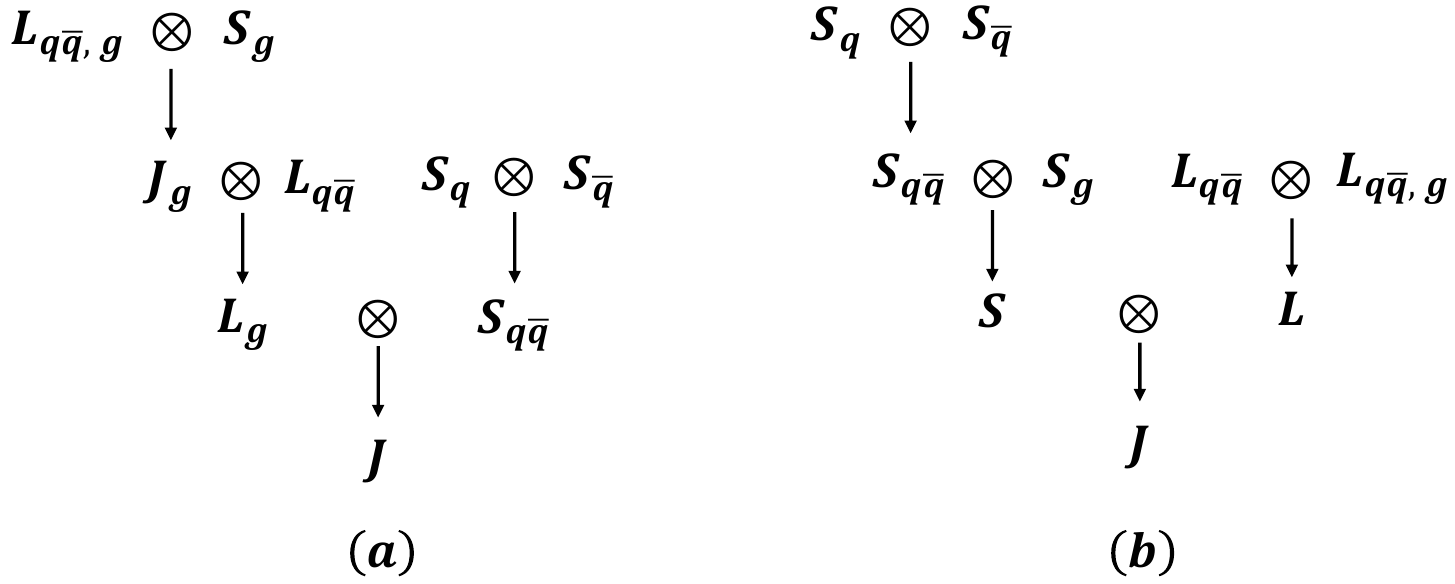}
    \caption{Two kinds of coupling schemes. The left panel shows the wave function coupling mode of Eq.~(\ref{j-j}), and the right panel shows the coupling mode of Eq.~(\ref{l-s}).}
    \label{coupling}
\end{figure}
In Refs.~\cite{Chen:2023ukh,Zhang:2025xee}, the coupling scheme of spin and orbit is a little bit like the "light degree of freedom" in heavy quark symmetry, i.e., the gluon field couples the two relative angular momenta $L_{q\bar{q},g}$ and $L_{q\bar{q}}$ first to become an excited constituent gluon, then this excited gluon couples to $S_{q\bar{q}}$ to obtain the total angular momentum $J$ (left panel of Fig.~\ref{coupling}), which can be symbolically written as
\begin{eqnarray}
    \psi_{q\bar{q}g}^{J,g} &=& \left[ \left[\psi_q^{\frac{1}{2}}\psi_{\bar{q}}^{\frac{1}{2}}\right]^{S_{q\bar{q}}} \left[ \left[\psi_{q\bar{q},g}^{L_{q\bar{q},g}} \psi_g^1\right]^{J_g} \psi_{q\bar{q}}^{L_{q\bar{q}}} \right]^{L_g} \right]^J.
    \label{j-j}
\end{eqnarray}
However, this kind of coupling scheme is a little complicated to be used into calculation since the spin wave functions of gluon and $q\bar{q}$ cluster are separated. In practice, to treat a three-body quantum system like baryon, the $L-S$ coupling scheme is often used (right panel of Fig.~\ref{coupling}), whose wave function can be constructed as
\begin{eqnarray}
    \psi_{q\bar{q}g}^{J, LS} &=& \left[\left[\left[\psi_q^{\frac{1}{2}}\psi_{\bar{q}}^{\frac{1}{2}}\right]^{S_{q\bar{q}}} \psi_{g}^{1}\right]^S \left[\psi_{q\bar{q}}^{L_{q\bar{q}}} \psi_{q\bar{q},g}^{L_{q\bar{q},g}}\right]^L\right]^J.
    \label{l-s}
\end{eqnarray}

Actually, from quantum theory of angular momentum \cite{Varshalovich:1988ifq}, we can build a relation between these two schemes, which can be explicitly written out by using a $6j$-symbol and a $9j$-symbol as
\begin{eqnarray}
    \psi_{q\bar{q}g}^{J,g} &=& \sum_{J_{q\bar{q}},S,L} \sqrt{(2J_{q\bar{q}}+1)(2L_g+1)}(-1)^{S_{q\bar{q}}+L_{q\bar{q}}+J_g+J}\nonumber\\
    &&\times\begin{Bmatrix}
    S_{q\bar{q}} & L_{q\bar{q}} & J_{q\bar{q}}\\
    J_g & J& L_g
    \end{Bmatrix}
    \sqrt{(2S+1)(2L+1)}\nonumber\\
    &&\times\sqrt{(2J_{q\bar{q}}+1)(2J_g+1)}\begin{Bmatrix}
    S_{q\bar{q}} & L_{q\bar{q}} & J_{q\bar{q}}\\
    1 & L_{q\bar{q},g} & J_g\\
    S & L & J
    \end{Bmatrix}\nonumber\\
    &&\times \psi^{J,LS}_{q\bar{q}g}.
\end{eqnarray}

\subsubsection{More details on the transition amplitude from a hybrid to two mesons}

Since our transition operator is written in momentum space, to adopt it to get the amplitude, we should firstly transform the wave functions into the momentum representation as
\begin{eqnarray}
    \psi_i^{L_i}(\boldsymbol{p}) &=& \sum_{n=1}^{n_{max}} (2\pi)^{\frac{3}{2}} c_n^{L_i} \frac{(-i)^{L_i}}{(2\nu_n)^{\frac{L_i}{2}+\frac{3}{4}}} \sqrt{\frac{2^{L_i+2}}{\sqrt{\pi}(2L_i+1)!!}}\nonumber\\
    &&\times e^{-\frac{p^2}{4\nu_n}}\mathcal{Y}_{L_i}(\vec{p}),
\end{eqnarray}
where $(2\pi)^{\frac{3}{2}}$ comes due to our notation that space vectors are normalized to $(2\pi)^3$.

Then, by a similar treatment that used in the $^3P_0$ model, for the process where a hybrid $A$ decays into two mesons $B$ and $C$, the transition amplitude, which is very similar as in Refs. \cite{Ding:2006ya}, is
\begin{widetext}
    \begin{eqnarray}
        M^{M_{J_A},M_{J_B},M_{J_C}} &=& \sum_{\renewcommand{\arraystretch}{.5}\begin{array}[t]{l}
        \scriptstyle M_{L_{q\bar{q},g}},M_{S_g},M_{L_{q \bar{q}}},M_{S_{q\bar{q}}},\\
        \scriptstyle M_{L_B},M_{S_B},M_{L_C},M_{S_C}\\
        \end{array}}\renewcommand{\arraystretch}{1}\!\!
        \mathcal{C} \mathcal{S} \left\langle L_{q\bar{q},g}, M_{L_{q\bar{q},g}} ; 1, M_{S_g} \mid J_g, M_{L_{q\bar{q},g}}+M_{S_g}\right\rangle\nonumber\\
        &&\times\left\langle L_{q \bar{q}}, M_{L_{q \bar{q}}} ; J_g, M_{L_{q\bar{q},g}}+M_{S_g} \mid L_{g}, M_{L_{q\bar{q},g}}+M_{S_g}+M_{L_{q \bar{q}}}\right\rangle \nonumber\\
        &&\times \left\langle L_{g}, M_{L_{q\bar{q},g}}+M_{S_g}+M_{L_{q \bar{q}}} ; S_{q \bar{q}}, M_{S_{q \bar{q}}} \mid J_A, M_{J_A}\right\rangle \nonumber\\
        &&\times \left\langle L_B, M_{L_B} ; S_B, M_{S_B} \mid J_B, M_{J_B}\right\rangle \left\langle L_C, M_{L_C} ; S_C, M_{S_C} \mid J_C, M_{J_C}\right\rangle  \nonumber\\
        &&\times[ \langle \phi_{B}^{14} \phi_{C}^{32}\mid\phi_{A}^{12}\phi_{0}^{34}\rangle \mathcal{I}^{(24)}+(-1)^{1+S_{q\bar{q}}+S_B+S_C}\langle \phi_{B}^{32} \phi_{C}^{14}\mid\phi_{A}^{12}\phi_{0}^{34}\rangle \mathcal{I}^{(13)}],
    \end{eqnarray}
    Here, $\mathcal{C}$, $\langle \phi_{B} \phi_{C}|\phi_{A}\phi_{0}\rangle$, and $\mathcal{S}$ represent color, flavour, and spin overlap factors, respectively. While $\mathcal{C}=\frac{2}{3}$ is a constant for hybrid decay, and $\mathcal{S}$ can be related to a $9-j$ symbol as
    \begin{eqnarray}
        \mathcal{S} &=& \sum_{S} \sqrt{6\left(2 S_B+1\right)\left(2 S_C+1\right)\left(2 S_{q \bar{q}}+1\right)}\left\{\begin{array}{ccc}
        \frac{1}{2} & \frac{1}{2} & S_B \\
        \frac{1}{2} & \frac{1}{2} & S_C \\
        S_{q \bar{q}} & 1 & S
        \end{array}\right\} \nonumber\\
        && \times\left\langle S_{q \bar{q}}, M_{S_{q \bar{q}}} ; 1, M_{s_g} \mid S, M_{S_B}+M_{S_C}\right\rangle\left\langle S_B, M_{S_B} ; S_C, M_{S_C} \mid S, M_{S_B}+M_{S_C}\right\rangle.
    \end{eqnarray}
    
    As for the momentum space integrals, they can be written in detail as
    \begin{eqnarray}
        \mathcal{I}^{(24)}(\mathbf{P,q,k}, m_1, m_2, m_4) &=& \int \frac{d^3 \mathbf{q} d^3 \mathbf{k}}{\sqrt{2 \omega_g}(2 \pi)^6} \psi_{L_{q\bar{q}} M_{L_{q\bar{q}}}}(\mathbf{P}+\mathbf{q}+\frac{m_1-m_2}{2m_1+2m_2}\mathbf{k}) \psi_{L_{q\bar{q},g} M_{L_{q\bar{q},g}}}(\mathbf{k})  \nonumber\\
        && \times \psi_{L_B M_{L_B}}^*(\frac{m_4}{m_1+m_4}\mathbf{P}+\mathbf{q}-\frac{\mathbf{k}}{2})  \psi_{L_C M_{L_C}}^*(\frac{m_4}{m_2+m_4}\mathbf{P}+\mathbf{q}+\frac{\mathbf{k}}{2}),
    \end{eqnarray}

    \begin{eqnarray}
        \mathcal{I}^{(13)}(\mathbf{P,q,k}, m_1, m_2, m_4) &=& \int \frac{d^3 \mathbf{q} d^3 \mathbf{k}}{\sqrt{2 \omega_g}(2 \pi)^6} \psi_{L_{q\bar{q}} M_{L_{q\bar{q}}}}(-\mathbf{P}+\mathbf{q}+\frac{m_1-m_2}{2m_1+2m_2}\mathbf{k}) \psi_{L_{q\bar{q},g} M_{L_{q\bar{q},g}}}(\mathbf{k})  \nonumber\\
        && \times \psi_{L_B M_{L_B}}^*(\frac{-m_4}{m_2+m_4}\mathbf{P}+\mathbf{q}+\frac{\mathbf{k}}{2})  \psi_{L_C M_{L_C}}^*(\frac{-m_4}{m_1+m_4}\mathbf{P}+\mathbf{q}-\frac{\mathbf{k}}{2}),
    \end{eqnarray}
     where $m_i$ is the mass of the constituent (anti-)quark $i$, $\omega_g$ is the energy of constituent gluon, $\mathbf{P}\equiv\mathbf{P}_B=-\mathbf{P}_C$ is the momentum of meson $B$ in the center of mass system of hybrid $A$, $\mathbf{q}$ is the relative momentum between the created quark-antiquark pair, and $\mathbf{k}$ is the relative momentum between constituent gluon and $q\bar{q}$ cluster in hybrid.
\end{widetext}

\subsubsection{Chiral constituent quark model}
When describing the interaction between $q\bar{q}$, the chiral quark model has become one of the most effective approaches to describe hadron spectra, hadron-hadron interactions and multiquark states.
The general form of multi-body Hamiltonian in the model is given as
\begin{align}
    H =& \sum_{i=1}^{n} \left( m_i + \frac{\mathbf{p_i}^2}{2m_i} \right) - T_{CM} \nonumber\\
    &+\sum_{j>i=1}^{n} \left[ V_{CON}(\mathbf{r}_{ij}) + V_{OGE}(\mathbf{r}_{ij}) + V_{GBE}(\mathbf{r}_{ij}) \right],
\end{align}
where $m_i$ is the constituent mass (quark, antiquark, or gluon), $\mathbf{p}_i$ is momentum of costituents, and $T_{CM}$ is the kinetic energy of the center-of mass.

Due to the fact that a nearly massless current light quark acquires a dynamical, momentum dependent mass (so-called constituent quark mass) for its interaction with the gluon medium, ChQM contains color confinement potential, one-gluon exchange potential (OGE) and Goldstone boson exchange potentials (GBE).
These three potentials reveal the most relevant features of QCD at low energy regime, i.e., confinement, asymptotic freedom and chiral symmetry spontaneous breaking.

Enough work in the past has done a good job investigating the states discovered experimentally, especially the multiquark candidates. Therefore, the interaction potentials between quarks and anti-quarks will adopt the form in previous work~\cite{Vijande:2004he,Yang:2011rp,Chen:2024ukv}. For color confinement potential, we consider central and spin-orbit contributions, if taking the screened form as an example, they can be written in detail as follows
\begin{align}
    V_{CON}^{C}(\mathbf{r}_{ij}) &= \left(-\boldsymbol{\lambda}_i^c \cdot \boldsymbol{\lambda}_j^{c*}\right) \left[-a_c(1-e^{-\mu_c r_{ij}}) + \Delta\right],  \\
    V_{CON}^{SO}(\mathbf{r}_{ij}) &= -\left(-\boldsymbol{\lambda}_i^c \cdot \boldsymbol{\lambda}_j^{c*}\right) \frac{a_c \mu_c e^{-\mu_c r_{ij}}}{4 m_i^2 m_j^2 r_{ij}} \left[ ((m_i^2 + m_j^2) \right.\nonumber \\
    &\quad\times (1 - 2a_s) + 4m_{i}m_{j}(1 - a_s)) (\boldsymbol{S}_+ \cdot \boldsymbol{L}) \nonumber \\
    &\quad\left. + (m_j^2 - m_i^2)(1 - 2a_s) (\boldsymbol{S}_- \cdot \boldsymbol{L}) \right], 
\end{align}
where $a_c$, $\mu_c$, and $\Delta$ are model parameters, and $\boldsymbol{\lambda}^c$ represent the SU(3) color Gell-Mann matrices.

One-gluon exchange potential contains coulomb, color-magnetism, spin-orbit, and tensor interactions, which arise from QCD perturbation effects at leading order as
\begin{align}
    V_{OGE}^{C}(\mathbf{r}_{ij}) =& \left(-\boldsymbol{\lambda}_i^c \cdot \boldsymbol{\lambda}_j^{c*}\right) \frac{\alpha_s}{4} \Big[ \frac{1}{r_{ij}} - \frac{1}{6m_im_j} \frac{e^{-r_{ij}/r_0(\mu)}}{r_{ij}r^2_0(\mu)}\nonumber\\
    &\times \left(\boldsymbol{\sigma}_i\cdot \boldsymbol{\sigma}_j\right)\Big],\\
    V_{OGE}^T(\mathbf{r}_{ij}) =& \left(-\boldsymbol{\lambda}_i^c \cdot \boldsymbol{\lambda}_j^{c*}\right) \frac{-\alpha_s}{16m_i m_j} \left[\frac{1}{r_{i j}^3}-\frac{\mathrm{e}^{-r_{i j} / r_g(\mu)}}{r_{i j}}\right.\nonumber\\
    &\times\left.\left(\frac{1}{r_{i j}^2}+\frac{1}{3 r_g^2(\mu)}+\frac{1}{r_{i j} r_g(\mu)}\right)\right] S_{i j},\\
    V_{OGE}^{SO}(\mathbf{r}_{ij}) =& \left(-\boldsymbol{\lambda}_i^c \cdot \boldsymbol{\lambda}_j^{c*}\right) \frac{-\alpha_s}{16 m_i^2 m_j^2} \left[\frac{1}{r_{i j}^3}-\frac{e^{-r_{i j} / r_g(\mu)}}{r_{i j}^3}\right.\nonumber\\
    &\times\left.\left(1+\frac{r_{i j}}{r_g(\mu)}\right)\right] \left[\left(\left(m_i+m_j\right)^2+2 m_i m_j\right) \right.\nonumber\\
    &\times\left(\boldsymbol{S}_{+} \cdot \boldsymbol{L}\right)+\left(m_j^2-m_i^2\right)\left(\boldsymbol{S}_{-} \cdot \boldsymbol{L}\right)\Big],
\end{align}
where $\mu$ is the reduced mass of two interacting quarks, $\boldsymbol{\sigma}$ represent the $SU(2)$ Pauli matrices, $r_0(\mu)=\hat{r}_0/\mu$, $r_g(\mu)=\hat{r}_g/\mu$, $\alpha_s$ denotes the effective scale-dependent strong running coupling constant of one-gluon exchange,
\begin{equation}
\alpha_{s} =\frac{\alpha_{0}}{\log\left(\frac{\mu^2+\mu_{0}^2}{\Lambda_{0}^2}\right)}.
\end{equation}

Due to chiral symmetry spontaneous breaking, Goldstone boson exchange potentials appear between light quarks ($u$, $d$ and $s$). Same as one-gluon-exchange, in this work, we also consider the center part in addition with tensor and spin-orbit contributions, with the tensor part arising from pseudoscalar meson exchange and the spin-orbit from scalar meson exchange, as
\begin{align}
    V_{GBE}\left({{\mathbf{r}}_{ij}}\right) &= V_{\sigma}({{\mathbf{r}}_{ij}}) + V_{\pi}({{\mathbf{r}}_{ij}})\sum_{a=1}^{3} \lambda_i^a \lambda_j^{a*} + V_{K}({{\mathbf{r}}_{ij}})\nonumber\\
    &\times\sum_{a=4}^{7}\lambda_i^a \lambda_j^{a*} + V_{\eta}({{\mathbf{r}}_{ij}})\left[\cos\theta_{P}(\lambda_i^8 \lambda_j^{8*})\right.\nonumber\\
    &-\left.\sin\theta_{P}(\lambda_i^0 \lambda_j^{0*})\right],
\end{align}
which can be written in detail as
\begin{align}
    V_{\chi=\pi,K,\eta}^{C}({{\mathbf{r}}_{ij}}) = & \frac{g^2_{ch}}{4\pi}\frac{m^2_\chi}{12m_im_j} \frac{\Lambda^2_\chi }{\Lambda^2_\chi-m^2_\chi} m_\chi\Bigg[ Y(m_{\chi}r_{ij}) \nonumber \\
    &\left. -\frac{\Lambda^3_\chi}{m^3_\chi} Y(\Lambda_{\chi}r_{ij}) \right] (\boldsymbol{\sigma}_i \cdot \boldsymbol{\sigma}_j), \\
    V_{\sigma}^C({{\mathbf{r}}_{ij}}) = & -\frac{g_{c h}^2}{4 \pi}\frac{\Lambda_\sigma^2}{\Lambda_\sigma^2-m_\sigma^2} m_\sigma\Bigg[Y\left(m_\sigma r_{i j}\right) \nonumber\\
    &\left.-\frac{\Lambda_\sigma}{m_\sigma} Y\left(\Lambda_\sigma r_{i j}\right)\right],\\
    V_{\chi=\pi,K,\eta}^T({{\mathbf{r}}_{ij}}) = & \frac{g_{\mathrm{ch}}^2}{4 \pi} \frac{m_\chi^2}{12 m_i m_j} \frac{\Lambda_\chi^2 }{\Lambda_\chi^2-m_\chi^2}m_\chi \Bigg[H\left(m_\chi r_{i j}\right)\nonumber\\
    &\left.-\frac{\Lambda_\chi^3}{m_\chi^3} H\left(\Lambda_\chi r_{i j}\right)\right]S_{i j} ,\\
    V_{\sigma}^{SO}({{\mathbf{r}}_{ij}}) =& -\frac{g_{c h}^2}{4 \pi} \frac{\Lambda_\sigma^2}{\Lambda_\sigma^2-m_\sigma^2} \frac{m_\sigma^2}{2 m_i m_j}\Bigg[G\left(m_\sigma r_{i j}\right)\nonumber\\
    &-\frac{\Lambda_\sigma^3}{m_\sigma^3} G\left(\Lambda_\sigma r_{i j}\right)\Bigg]\left(\boldsymbol{S}_+\cdot\boldsymbol{L}\right).
\end{align}
Here, $Y(x)$ is the standard Yukawa function, $H(x)=(1+3/x+3/x^2)Y(x)$, $G(x)=(1+1/x)Y(x)/x$, $S_{ij}=\boldsymbol{\sigma}_i\cdot\boldsymbol{\sigma}_j-\frac{3(\boldsymbol{\sigma}_i\cdot\boldsymbol{r}_{ij})(\boldsymbol{\sigma}_j\cdot\boldsymbol{r}_{ij})}{\boldsymbol{r}_{ij}^2}$ is the tensor operator, $\lambda^a$ are the Gell-Mann matrices; $\Lambda_s$ is the cut-off of meson $s$, $m_{\chi=\pi,K,\eta}$ are the masses of Goldstone bosons, $m_{\sigma}$ is determined through $m_\sigma^{2} = m_{\pi}^{2}+4m_{u,d}^{2}$, and $g^2_{ch}$ is the chiral field coupling constant,
which is determined from the $NN\pi$ coupling constant through
\begin{equation}
\frac{g^2_{ch}}{4\pi}=\frac{9}{25}\frac{g^2_{\pi NN}}{4\pi}\frac{m^2_{u,d}}{m^2_N}.
\end{equation}

\subsection{More details on numerical calculation}

We keep the same model parameters as the previous works on meson spectra calculations. Since we adopt three kinds of confinements, i.e., screened form \cite{Vijande:2004he}, linear form \cite{Yang:2011rp}, and square form \cite{Chen:2024ukv}, there are three corresponding groups of model parameters, which are extracted from Refs.~\cite{Vijande:2004he,Yang:2011rp,Chen:2024ukv} into Table~\ref{tab:parameter}.

\begin{table}[!h]
    \centering
    \caption{Parameters of the chiral quark model with three confinement potential forms: screened(n=0) \cite{Vijande:2004he}, linear(n=1) \cite{Yang:2011rp}, square(n=2) \cite{Chen:2024ukv}. The Goldstone-boson exchange interaction parameters are consistent. Masses of $\pi$, $\eta$, $K$ adopt experimental values, while other parameters --- $m_\sigma$ = 3.42 $\rm fm^{-1}$, $\Lambda_\pi = \Lambda_\sigma$ = 4.2 $\rm fm^{-1}$, $\Lambda_\eta = \Lambda_K$ = 5.2 $\rm fm^{-1}$, $\theta_{p} = -15^\circ$, $g_{ch}^2/(4\pi)$ = 0.54 --- are adopted from Ref.~\cite{Vijande:2004he}}\label{tab:parameter}
    \renewcommand\arraystretch{1.5}
    \setlength{\tabcolsep}{1.5mm}     
    \begin{tabular}{cccc} \hline\hline   
    & & & [~~Scr.~,~~Lin.~~,~Squ.~~]\\\hline 

    Gluon~mass & $m_{g}$~(MeV) & & 450\\\hline
    
    Quark~masses & $m_{u,d}$~(MeV) & & 313\\
    
    & $m_{s}$~(MeV) & & [~~555~~,~~525~~,~~536~~]\\
    
    & $m_{c}$~(MeV) & & [~1752~,~~1731~,~1728~]\\
        
    & $m_{b}$~(MeV) & & [~5100~,~~5100~,~5112~]\\
    \hline
    
    Confinement & $a_{c}$~$\rm (MeV~fm^{-n})$ & & [~~430~~,~~160~~,~~101~~]\\
    
    & $\Delta$~(MeV) & & [~181.1,~-131.1,~-78.3~]\\
    
    & $\mu _{c}$~$\rm (fm^{-1})$ & & [~~0.7~~,~~~~-~~~~,~~~~-~~~]\\
    
    & $a _{s} $ & & 0.777\\
    
    \hline
    
    OGE & $\alpha_{0}$ & & [~2.12~,~~2.65~~,~3.67~]\\
    
    & $\Lambda_{0}$~$\rm (fm^{-1})$ & & [~0.113,~~0.075,~0.033]\\
    
    & $\mu_{0}$~(MeV) & & 36.976\\
    
    & $\hat{r}_{0}$~(MeV~fm) & & 28.17\\
    
    & $\hat{r}_{g}$~(MeV~fm) & & 34.5\\
    
    \hline\hline
    \end{tabular}
\end{table}

By using these three groups of parameters, we calculate the low-lying light meson spectra, and the results are collected into Table~\ref{tab:meson}. The results show that the spectra of low-lying isospin-1 and isospin-$\frac{1}{2}$ mesons can be nicely reproduced, while the spectra of isospin-0 mesons are not so good. That is because all these light isospin-0 mesons are actually a mixture of $q\bar{q}$ and $s\bar{s}$ states, i.e., $|meson\rangle_{I=0} = \cos\theta|q\bar{q}\rangle_{I=0}+\sin\theta|s\bar{s}\rangle_{I=0}$, while in our calculation we treat them all as pure $q\bar{q}$ or $s\bar{s}$ states. However, we want to emphasize here that in our practical calculation on the decay width, we use the wave function of the physical states, i.e., $|meson\rangle_{I=0}$, and the corresponding mixing angles $\theta$ are kept the same with our previous work \cite{Zhang:2025xee}.

\begin{table}[!h]
    \centering
    \caption{Meson spectrum calculated by using three sets of parameters and confinement potentials (unit: MeV). }\label{tab:meson}
    \renewcommand\arraystretch{1.5}
    \setlength{\tabcolsep}{2.0mm}
    \begin{tabular}{ccccc} \hline\hline
    States & & [~~~Scr.~~~,~~~Lin.~~~~,~~~Squ.~~~] & & Exp.\\\hline
    
    $\pi$ & & [~~132.18~,~~140.08~~,~134.87~~] & & 139.57\\
    $\eta$ & & [~~684.74~,~~680.19~~,~669.21~~] & & 547.86\\
    $\rho$ & & [~~773.92~,~~775.33~~,~772.26~~] & & 775.26\\
    $\omega$ & & [~~697.89~,~~703.70~~,~701.59~~] & & 782.66\\
    $K$ & & [~~472.58~,~~496.21~~,~489.37~~] & & 493.68\\
    $K^*$ & & [~~908.39~,~~917.90~~,~913.55~~] & & 891.67\\
    $\eta^\prime$ & & [~~824.19~,~~832.80~~,~821.47~~] & & 957.78\\
    $h_1$(1170) & & [~1247.02~,~1271.19~,~1314.82~] & & 1166.00\\
    $b_1$(1235) & & [~1234.79~,~1250.19~,~1281.93~] & & 1229.50\\
    $a_1$(1260) & & [~1204.76~,~1213.91~,~1238.72~] & & 1230.00\\
    $f_1$(1285) & & [~1149.08~,~1212.92~,~1283.19~] & & 1281.80\\
    $\pi(1300)$ & & [~1286.99~,~1345.44~,~1453.58~] & & 1300.00\\
    $K_1(1270)$ & & [~1342.73~,~1293.69~,~1287.15~] & & 1253.00\\
    $K_1(1400)$ & & [~1416.25~,~1425.47~,~1453.16~] & & 1403.00\\
    $K(1460)$ & & [~1465.49~,~1490.60~,~1571.79~] & & 1482.40\\\hline\hline   
    \end{tabular}
\end{table}

Finally, the detailed numerical results of the low-lying $1^{-+}$ light hybrids are presented in Table~\ref{qqgmass}, where for $q\bar{q}g(1)$, i.e., the $\pi_1$ family, the first line is the ground state, the second line is mainly due to the radial excitation between constituent gluon and $q\bar{q}$ cluster, and the third line is mainly caused by the radial excitation between quark and antiquark. In addition, due to the different behaviors of confinements, uncertainties on the first excited state are a little large, while the difference on the ground state is very small, which is consistent with our previous experience on meson and baryon calculations that different kinds of confinements affect little on the mass of ground states.

\begin{table}[!h]
    \centering
    \caption{Predicted masses of $1^{-+}$ hybrid mesons under three sets of confinement potentials (unit: MeV).} \label{qqgmass}
    \renewcommand\arraystretch{1.5}
    \setlength{\tabcolsep}{5.5mm}
    \begin{tabular}{cccc} \hline\hline
    States(I)  & $\rm M_{Scr.}$ & $\rm M_{Lin.}$ & $\rm M_{Squ.}$\\\hline
    \multirow{3}{*}{$q\bar{q}g$(1)}  & 1608.4 & 1631.3 & 1652.4\\\cline{2-4} 
      & 1890.2 & 1987.1 & 2130.1\\\cline{2-4} 
      & 2087.0 & 2294.5 & 2600.5\\\hline 
    $q\bar{q}g$(0)  & 1596.2 & 1614.7 & 1627.2\\\hline 
    $s\bar{s}g$(0)  & 1974.6 & 1927.9 & 1957.8\\\hline 
    $q\bar{s}g$($\frac{1}{2}$)  & 1793.9 & 1785.2 & 1830.9\\\hline\hline
    
    \end{tabular}
\end{table}

\subsection{Impact of constituent gluon mass on the spectra of hybrid}

Actually, as we have said in the introduction, there are different constituent gluon masses given by different models. As for their values, we collect them into the Table~\ref{gluonmass} as follows.
\begin{table}[ht]
    \centering
    \caption{Comparison of the gluon mass estimates from various methods. The prediction in this work is obtained by synthesizing the gluon mass range from the above methods and calculating the corresponding hybrid mass~(unit: GeV).} \label{gluonmass}
    \renewcommand\arraystretch{1.5}
    \setlength{\tabcolsep}{1.5mm}
    \begin{tabular}{ccccc} \hline\hline
    \multicolumn{2}{c}{Method} & Gluon~Mass &~& Light $1^{-+}~q\bar{q}g$ \\\hline
    \multicolumn{2}{c}{\multirow{2}{*}{\shortstack{Dyson-Schwinger equations \\ \cite{Cornwall:1981zr,Halzen:1992vd,Binosi:2022djx}}}} & \multirow{2}{*}{0.3-0.7} & & \multirow{2}{*}{-} \\ 
    &\\
    \multicolumn{2}{c}{Lattice QCD~\cite{Bernard:1981pg,MILC:1997usn,Bernard:2003jd,Lacock:1996ny,Mei:2002ip,McNeile:1998cp}} & 0.5-0.8 & & 1.6-2.0 \\\hline
    \multicolumn{2}{c}{Bag model~\cite{Barnes:1982zs,Barnes:1982tx,Donoghue:1983fy}} & 0.5-1.0 & & 1.2-1.7  \\ 
    
    \multicolumn{2}{c}{\multirow{2}{*}{\shortstack{Constituent gluon model \\ \cite{Iddir:2007dq,Szczepaniak:2001rg,Ishida:1991mx,Horn:1977rq}}}} & \multirow{2}{*}{0.5-0.8} & & \multirow{2}{*}{1.5-2.1} \\ 
    &\\ 
    \hline\hline 
      
    \end{tabular}
\end{table}
Thus, considering the conclusions given by Ref.~\cite{Binosi:2022djx} that $m_g \approx \frac{1}{2}m_p$ and Table~\ref{gluonmass}, we carry on a study on the impact of constituent gluon mass on the spectra of hybrid, where the constituent gluon mass is taken from 0.4~GeV to 0.8~GeV, and the result is collected in the Table~\ref{qqgmassrange}.
\begin{table}[ht]
    \centering
    \caption{Mass positions of ground $1^{-+}$ hybrid mesons for gluon mass around 0.4-0.8 GeV (unit: GeV).} \label{qqgmassrange}
    \renewcommand\arraystretch{1.5}
    \setlength{\tabcolsep}{6.5mm}
    \begin{tabular}{ccc} \hline\hline
    Gluon~Mass  & States(I) & Mass \\\hline
    \multirow{4}{*}{0.4-0.8}  & $q\bar{q}g$(1) & 1.58-1.90 \\\cline{2-3} 
      & $q\bar{q}g$(0) & 1.57-1.88 \\\cline{2-3} 
      & $s\bar{s}g$(0) & 1.91-2.19 \\\cline{2-3} 
      & $q\bar{s}g$($\frac{1}{2}$) & 1.75-2.02\\
    \hline\hline
    \end{tabular}
\end{table}

\begin{figure}[!h]
    \centering
    \includegraphics[width=0.48\textwidth]{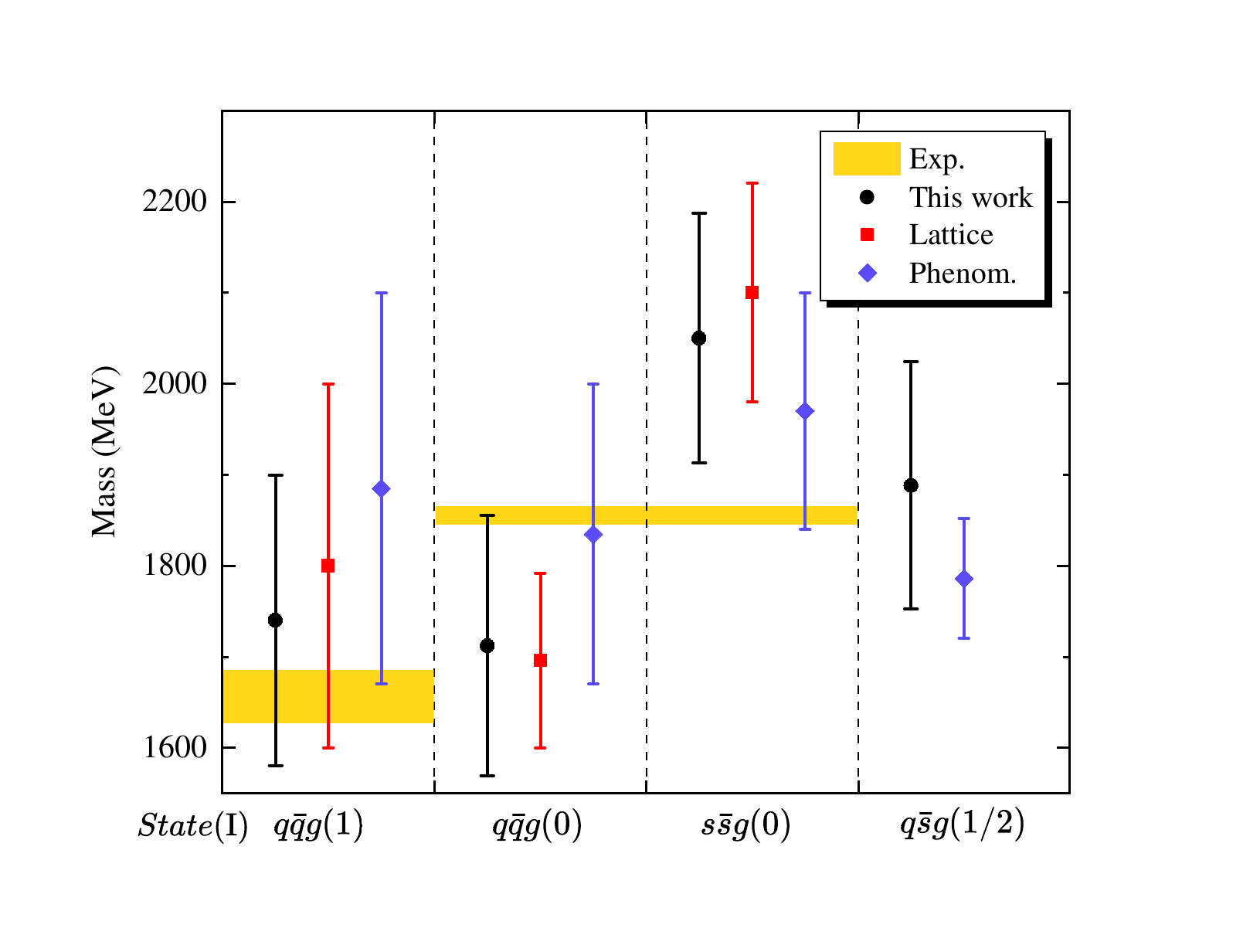}
    \caption{Impact of different constituent gluon masses on the  spectra of ground hybrid states with quantum number $1^{-+}$, where states are distinguished by compositeness and isospin, and the error bar line means a collection of different results obtained by different models or works. Here, the black lines denote our results, where the constituent gluon mass varies from 0.4~GeV to 0.8~GeV, the yellow bands reflect experimental data from Ref.~\cite{ParticleDataGroup:2024cfk}, the red lines correspond to lattice QCD results referenced in \cite{Bernard:1981pg,Lacock:1996ny,MILC:1997usn,McNeile:1998cp,Mei:2002ip,Bernard:2003jd,Dudek:2013yja}, and the blue lines represent results from other phenomenological models \cite{Meyer:2015eta,Shastry:2022mhk,Barnes:1982zs,Barnes:1982tx,Donoghue:1983fy,Iddir:2007dq,Szczepaniak:2001rg,Ishida:1991mx,Horn:1977rq,Tan:2024grd}.} 
    \label{fig:qqgg}
\end{figure}

From Table~\ref{qqgmassrange}, we can see that although the constituent gluon mass changes 400 MeV, the masses of ground $1^{-+}$ hybrid mesons change only around 200-300 MeV, this is mainly because the coupling constant between quark and gluon, i.e., $\alpha_s$, is non-linearly decreased with linearly increased reduced mass. In addition, from a comparison between Table~\ref{qqgmassrange} and Table~\ref{gluonmass} (or Fig.~\ref{fig:qqgg}), we can see that our results still match the conclusions given by other models, which may prove our opinion in this work that proper constituent gluon mass is possibly the final piece to construct the hybrid meson spectra.

\end{document}